\documentclass[a4paper,11pt]{article}
\usepackage{subcaption, hyperref} 
\usepackage{comment}
\usepackage{physics}
\usepackage{overpic}
\usepackage{graphicx,amsfonts,amssymb,amsmath,amsthm,bbold,dsfont}
\usepackage{xcolor,cite}
\usepackage{braket,enumerate}
\usepackage{mathtools}
\newcommand{\eps}{\varepsilon}
\newcommand{\E}{{ \mathbb{E}}}

\newcommand{\M}{{\mathcal{M}}}
\newcommand{\cN}{{\mathcal{N}}}

\newtheorem{thm}{Theorem}[section]

\newtheorem{assumption}[thm]{Assumption}

\newtheorem{pr}[thm]{Proposition}

\newtheorem{definition}[thm]{Definition}
\theoremstyle{remark}

\newtheorem{example}[thm]{Example}

\newtheorem{remark}[thm]{Remark}

\newtheorem{tab}{Table}

\newcommand{\nab}{{\mathcal{N}_{A\to B}}}

\newcommand{\W}{{\mathcal{W}}}
\newcommand{\caln}{{\mathcal{N}}}

\newcommand{\Renyi}{R{\'e}nyi~}

\author{Yonglong Li, Christoph Hirche, and Marco Tomamichel\thanks{Y.~Li and M.~Tomamichel are with the Department of Electrical and Computer Engineering, National University of Singapore (NUS). C.~Hirche is  with Zentrum Mathematik, Technical University of Munich. C.~Hirche and M.~Tomamichel are also with the Center for Quantum Technologies (CQT), NUS.  (e-mails: \url{{elelong, c.hirche, marco.tomamichel}@nus.edu.sg}).} }
\begin{document}
\date{}
\title{Sequential Quantum Channel Discrimination}
\maketitle
\begin{abstract}
We consider the {\em sequential} quantum channel discrimination problem using adaptive and non-adaptive strategies. In this setting the number of uses of the underlying quantum channel is not fixed but a random variable that is either bounded in expectation or with high probability. We show that both types of error probabilities decrease to zero exponentially fast and, when using adaptive strategies, the rates are characterized by the measured relative entropy between two quantum channels, yielding a strictly larger region than that achievable by non-adaptive strategies. Allowing for quantum memory, we see that the optimal rates are given by the regularized channel relative entropy. Finally, we discuss achievable rates when allowing for repeated measurements via quantum instruments and conjecture that the achievable rate region is not larger than that achievable with POVMs by connecting the result to the strong converse for the quantum channel Stein's Lemma.
\end{abstract}
\section{Introduction}\label{sec:introduction}
Quantum hypothesis testing between different quantum sources is of central importance to a variety of quantum information processing tasks. In particular, discrimination of two quantum states has long been an active research area in quantum information theory. Here, the goal is to find tests that give the optimal trade-off between two kinds of error probabilities, namely the probabilities of false detection and false rejection. A typical setting is to consider an asymptotic scenario where an infinite number of copies of the state are available. For state discrimination this is a well explored problem~\cite{audenaert07-3,Petz1991,Nagaoka2000,Hayashi2004,Nagaoka2006,Hayashi2007,Audenaert2007,Nussbaum2009,berta2021composite}. 

Sequential methods for the classical hypothesis testing problem were first proposed in~\cite{Wald1945} and have later been expanded into a subject called sequential analysis. The key merit of sequential analysis is that the number of samples used in the statistical procedure is not fixed in advance of the statistical experiment and, given the tolerance error, the {\em average} number of samples needed is much less than that in the fixed-sample statistical experiment. In recent work, sequential hypothesis testing was extended to the setting of quantum states~\cite{quantumSHT,li2021optimal}. 

Another problem with considerable recent progress is that of quantum channel discrimination, which is a natural extension of the state version~\cite{Wilde2020,berta2019stein,wang2019resource,fang2020chain}. Most notably, these works determined the optimal asymptotic rate in asymmetric quantum channel discrimination which is known as a quantum Stein's Lemma. 

In this work, we combine these two fields and discuss sequential hypothesis testing between two quantum channels, determining the optimal rate regions under certain expectation and probabilistic constraints. We consider several different strategies including non-adaptive and adaptive strategies, and adaptive strategies with quantum memory. Ultimately showing that under the last set of strategies both errors decay exponentially at a rate given by the regularized channel relative entropy. Finally, we give achievable and converse bounds in the most general setting of strategies using quantum instruments potentially measuring the same states multiple times. The latter bounds match conditional on a conjecture that is related to the strong converse of the quantum Stein's lemma for quantum channels.

\paragraph*{Previous Results.}
Many of the early results on quantum state discrimination are reviewed in~\cite{audenaert07-3}. The generalization of Stein's lemma~\cite{Petz1991,Nagaoka2000} for quantum hypothesis testing establishes that the error of the second kind decays exponentially with the Stein's exponent given by the \emph{quantum relative entropy} when the first kind of error is upper bounded by a given constant. On the other hand, if both errors decrease exponentially, the optimal trade-off between the decay rates is governed by the quantum Hoeffding bound~\cite{Hayashi2004,Nagaoka2006,Hayashi2007}. In the Bayesian case, that is, the quantum state is prepared according to some prior probability mass function, the total error probability decreases to zero exponentially fast with exponent governed by the quantum Chernoff exponent~\cite{Audenaert2007,Nussbaum2009}. Beyond this, second-order refinements to the Stein's exponent were derived in~\cite{tomamichel12, li12} and the moderate deviation regime where one error probability decreases sub-exponentially has been analyzed in~\cite{chubb17,cheng17}.

 In~\cite{Hayashi2009}, Hayashi studied the classical channel discrimination problem and showed that adaptive protocols do not improve the error exponents in the Stein, Chernoff and Hoeffding regimes. In~\cite{Wilde2016}, the authors studied the discrimination of an arbitrary quantum channel and a ``replacer'' channel and showed that adaptive strategies provide no advantage over non-adaptive tensor-power strategies asymptotically. In~\cite{Wilde2020}, the authors introduced the amortized quantum channel divergence between quantum channels and showed that it is a general converse bound for the Stein's exponent. For several classes of channels~\cite{Wilde2020} showed that the error exponents of adaptive protocols are the same as those obtained by using non-adaptive protocols. In particular this applies to classical-quantum channels, see also~\cite{berta2019stein}. In~\cite{wang2019resource} it was subsequently shown that the amortized channel relative entropy is indeed also  achievable using adaptive strategies. Finally,~\cite{fang2020chain} showed that the amortized channel relative entropy is equal to the regularized channel relative entropy, which in turn is achievable by parallel strategies with quantum memory. This means that adaptive strategies are not more powerful than parallel strategies in this setting. 
In~\cite{Fawzi2021}, the authors introduces a new divergence using the weighed geometric mean between two operators and derived the strong converse exponent for quantum channel discrimination. 

For classical hypothesis testing problems between two probability distributions $P_0$ and $P_1$, when the expected number of samples is bounded by $n$, it was shown in~\cite{WaldWolf} that   there exists a sequence of tests---namely sequential probability ratio tests (SPRTs)---such that the exponents of the errors of the first and second kind {\em simultaneously} assume the extremal values  $D(P_1\| P_0)$ and $D(P_0\| P_1)$. This significantly improves the classical Hoeffding bound of the error exponents~\cite{Hoeffding1965, Blahut} where if one error exponent assumes its extremal value---the relative entropy---the other necessarily vanishes. The sequential approach to quantum hypothesis testing was first explored in~\cite{slussarenko17}. In a recent paper~\cite{quantumSHT},  it was shown that {\em sequential} quantum hypothesis test can reduce the number of samples needed compared with the fixed-length quantum hypothesis test. Also in~\cite{quantumSHT}, a converse was shown, that is, the expectation of the number of quantum states in the sequential procedure was lower bounded by a function of the tolerance error probabilities. In~\cite{li2021optimal}, the authors considered the sequential hypothesis testing of two quantum states under a different type of constraint on the number of states used
 and proposed an adaptive strategy to achieve the lower bound given in~\cite{quantumSHT}. Therefore,~\cite{li2021optimal} characterized the regions of all achievable error exponents of the two kinds of error probabilities.

 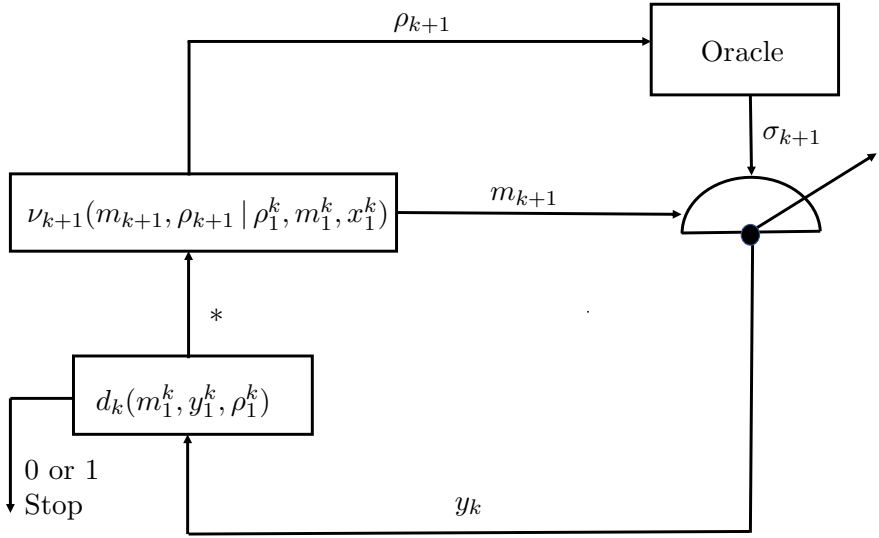
\begin{figure}
\centering
\begin{overpic}[width=.7\textwidth]{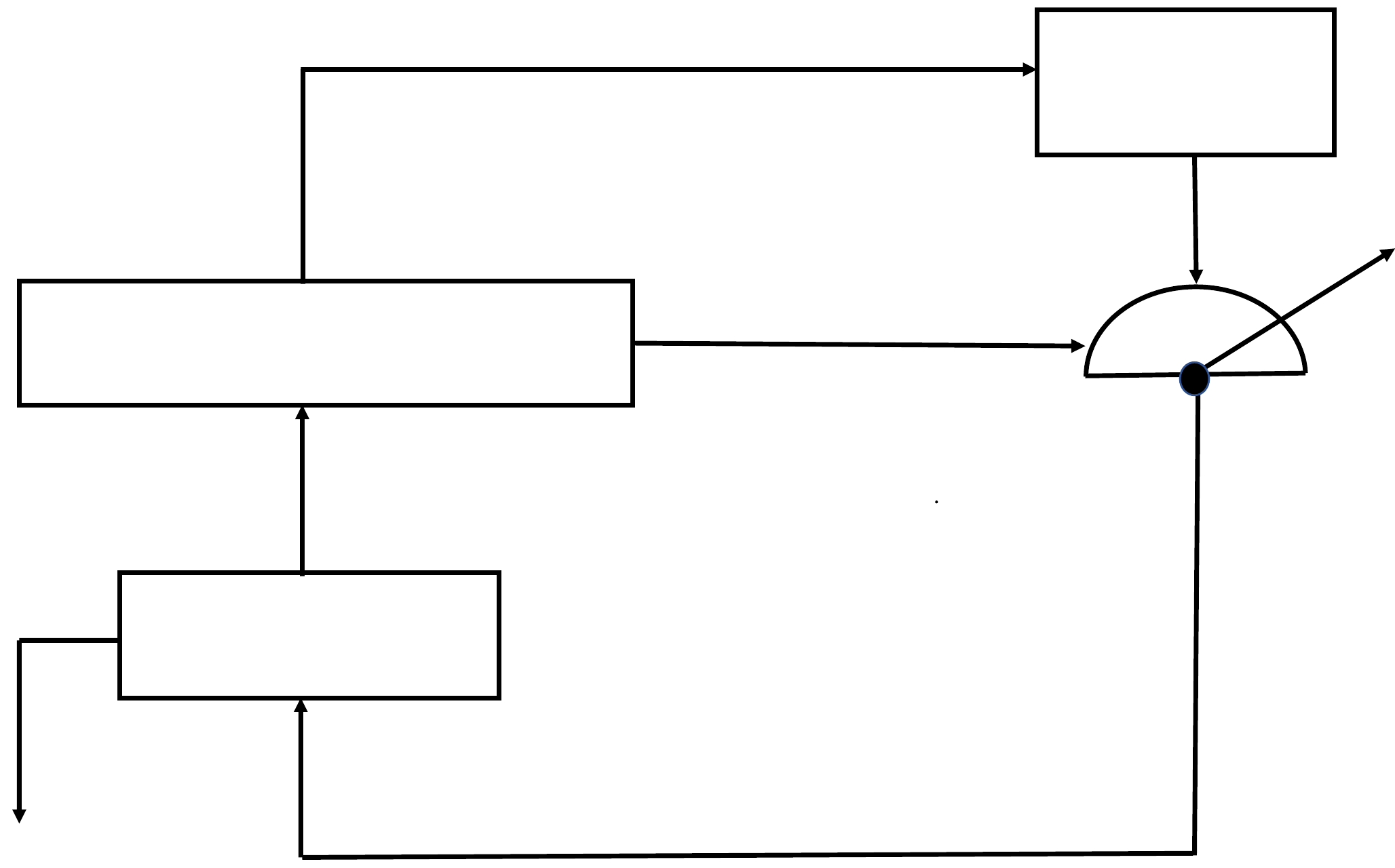}
\put(56,39){$m_{k+1}$}
\put(80,55){Oracle}
\put(87,46){$\sigma_{k+1}$}
\put(2.2,36.5){ $\nu_{k+1}(m_{k+1},\rho_{k+1}\,|\,\rho_1^{k},m_1^k,x_1^k)$}
\put(24,25){$*$}
\put(10,15){ $d_{k}(m_1^k,y_1^k,\rho_1^k)$}
\put(52,3.5){$y_k$}
\put(3,3){Stop}
\put(3,7){$0$ or $1$}
\put(45,59){$\rho_{k+1}$}
\end{overpic}
\caption{The structure of a general adaptive sequential channel discrimination protocol without quantum memory. At step $k$, the measurement outcome $y_k$ is considered together with all previous measurement results $y_{1}^{k-1}$ and all previous choices of the agent, $m_{1}^k$ and $\rho_1^k$. The decision function $d_k$ either decides to stop (outputting the hypothesis $0$ or $1$) or continue. In the latter case a new input state $\rho_{k+1}$ and a measurement $m_{k+1}$ are sampled according to the distribution $\nu_{k+1}$.
Finally, the channel oracle is called with input $\rho_{k+1}$, producing the output state $\sigma_{k+1}$ that will be measured using $m_{k+1}$.}
\label{fig:adaptive}
\end{figure}
 
\paragraph*{Outline.} 
The paper is structured as follows. In Section~\ref{sec:problemformulation}, we will introduce the notation used throughout the paper and the mathematical formulation of the problem. In Section~\ref{sec:mainresults}, the main results and the corresponding proofs are presented.

\section{Problem Formulation}\label{sec:problemformulation}

\subsection{Notation}
In this work, we consider finite-dimensional quantum systems. Throughout the paper, $A$, $B$, $C$, etc, denote quantum systems, but also the corresponding finite-dimensional Hilbert spaces. With $|A|$, $|B|$, $|C|$, etc, we denote the dimensions of the corresponding quantum systems.  Let $\mathcal{L}(A)$ be the set of all linear operators from $A$ to $A$. A quantum channel $\nab$ is a completely positive trace-preserving linear map from $\mathcal{L}(A)$ to $\mathcal{L}(B)$ (for more details on quantum channels, see~\cite[Chapter 5]{HayashiQIT}). Let $\mathcal{D}_{RA}$ be the set of bipartite quantum states over the quantum system $RA$. Let $\mathcal{Y}$ be some finite alphabet. A set $m=\{m_y:y\in\mathcal{Y}\}$ of  $|A|\times |A|$ positive-definite matrices is a positive operator valued measure (POVM) if $\sum_{y\in\mathcal{Y}}m_y=\mathbb{1}_A$. A POVM $m=\{m_y:y\in\mathcal{Y}\}$ is called a \emph{projector valued measure} (PVM) if each $m_y$ is a projector, that is, $m_y^2=m_y$. Let $\mathcal{M}_{\mathcal{Y}}$ be the set of POVMs with outcomes in $\mathcal{Y}$.


\subsection{The Sequential Quantum Channel Discrimination Problem}\label{sec:model}
We consider the following binary channel discrimination problem: 
$$H_0:\ \mathcal{W}=\cN_{0,A\to B}\quad H_1:\ \mathcal{W}=\cN_{1,A\to B},$$
where $\cN_{0,A\to B}$ and $\cN_{1,A\to B}$ are two quantum channels from $A$ to $B$. Throughout the rest of the paper, $\cN_{\nu,A\to B}$ will be denoted as $\cN_{\nu}$ for $\nu\in\{0,1\}$.


Let $\{R_i\}_{i=1}^{n}$ be a sequence of ancilla systems, each of which is an identical copy of some finite dimensional system $R$. Let $\mathcal{Y}$ be a finite alphabet and let $\mathcal{M}_{\mathcal{Y}}$ be the set of POVMs whose elements are of dimension $|A||B|\times |A||B|$. Throughout the paper $M$ is used to denote a random POVM and $m$ is used to be a realization of $M$.  For each $k\in\mathbb{N}$, let $d_k$ be a function from $\big(\M_{\mathcal{Y}}\times\mathcal{D}_{RA}\times\mathcal{Y}\big)^{k}$ to $\{0,1,*\}$. Intuitively, $d_k$ is the {\em decision function} at time $k$: based on all the input states, POVMs and outcomes before time $k+1$, $d_k\not=*$ means the testing procedure stops before time 
$k+1$ and the experimenter makes the decision that $H_{d_k}$ is true; $d_k=*$ means the testing procedure continues. At time $1$, the experimenter prepares the quantum state and the POVM $(\rho_1^{R_1A_1},M_1)$ randomly according to some probability measure $\nu_1$, and then passes the state through the underlying quantum channel $\W$ and obtains the output state $\sigma_1^{RB}=\mathrm{id}_{R_1}\otimes \W(\rho_1^{R_1A})$. Then the experimenter applies the POVM $M_1$  to $\sigma_1^{R_1B_1}$ obtaining the outcome $Y_1$. Then the experimenter applies a decision function $d_1(\rho_1^{R_1A_1},M_1,Y_1)$ to decide whether to accept $H_0$ or $H_1$ or to continue the experiment. If $d_1=*$, then the experimenter chooses to continue the experiment and prepares a new quantum state and a new POVM $(\rho_2^{R_2A_2},M_2)$ according to some conditional probability measure $\nu_2(\mathrm{d}\rho,\mathrm{d}m_2|\rho_1^{R_1A_1},M_1,Y_1)$ and obtains the output state $\sigma_{2}^{R_2B_2}$ by passing $\rho_2^{R_2A_2}$ through the underlying quantum channel. Then the experimenter applies some random POVM $M_2$ to $\sigma_2^{R_2B_2}$ and obtains the outcome $Y_2$. Based on $(\rho_1^{R_1A_1},\rho_2^{R_2A_2},M_1^2,Y_1^2)$, the experimenter applies some decision function $d_2$ to decide whether to accept one of the hypothesis or continue the experiment. This process continues until the experimenter accepts one of the hypothesis. Then $\{(M_i,\rho_i^{R_iA_i},Y_i)\}_{i=1}^{\infty}$ is a sequence of random variables taking values in $\M_{\mathcal{Y}}\times\mathcal{D}_{RA}\times\mathcal{Y}$. For notational convenience, $\rho_k^{RA}$ will be abbreviated as $\rho_k$.  The joint conditional density function of $(M_1^k,\rho_1^k,Y_1^k)$ is 
\begin{align}
&\hspace{-0.5cm}\mu^{(k)}(m_1^k,\rho_1^k,y_1^{k})=\prod_{j=1}^{k}\left\{\nu_j(\mathrm{d}\rho,\mathrm{d}m|x_1^{j-1},m_1^{j-1})\times\Tr[\mathcal{W}(\rho_j)m_{j}(x_j)]\right\},
\end{align}
for any $y_1^k$, $\rho_1^k$, and $m_1^k$. The described strategies are inherently {\em adaptive} as the input state and the measurement at each time depends on previous input states, measurements and outcomes of the measurements. When $\{\nu_j\}_{j=1}^{\infty}$ are probability measures on $\M_{\mathcal{Y}}\times\mathcal{D}_{RA}\times\mathcal{Y}$, the strategies are {\em non-adaptive} as the input state and the POVM chosen each time do not depend on the choices of previous rounds. The first time $k$ with $d_k\not=*$ is the number of uses of the underlying quantum channel and is denoted by $T$. The stopping time $T$ is well defined with respect to the filtration generated by $\{(M_j,\rho_j,Y_j)\}_{j=1}^{\infty}$.

\renewcommand{\P}{\mathbb{P}}
We call $\mathcal{S}=(\{\nu_k,d_k\}_{k=1}^{\infty},T)$ a {\em sequential quantum channel discrimination strategy}. In the following we study sequences of sequential quantum channel discrimination strategies $\mathcal{S}_n$, indexed by $n \in \mathbb{N}$. To simplify notation we use
$\P_{n,i}$ to denote $\P_{\mathcal{S}_n, \caln_i}$ for $i \in \{0,1\}$. The notation $\E_{n,i}[\cdot]$ means that the expectation is taken with respect to the probability measure $\P_{n,i}$.
We consider two types of constraints on the number of states $T_{n}$ used during the test. The first type of constraint is the {\em expectation constraint}: 
$
\max_{i \in \{0,1\}}\E_{n,i}[T_{n}]\le n.
$ In other words, the average number of copies used in the testing procedure should be bounded by some  number $n$. The second type of constraint is the {\em probabilistic constraint}~\cite{anusha,litan}
$
\max_{i \in \{0,1\}}\P_{n,i}(T_{n}>n)<\eps
$ for some fixed $\eps \in (0,1)$. In other words, the number of copies of the state used during the testing procedure should be bounded by some  number $n$ with probability larger than $1-\eps$.

We study the trade-off between the error probabilities $(\alpha_n,\beta_n)$ and either the expectation or the probabilistic constraint on the number of copies of the state used during the test procedure. The first type of error is quantified by the probability that the experimenter declares that hypothesis $1$ is in effect when, in fact, hypothesis $0$ is true, i.e., 
$
\alpha_n := \P_{0,n}(d_{T_{n}} = 1). 
$
On the other hand, the second type of error probability is 
$
\beta_n := \P_{n,1}(d_{T_{n}} = 0). 
$

\begin{definition}[Achievable Error Exponent Pairs] \label{def:ach_ee}
A pair $(R_0, R_1) \in\mathbb{R}_+^2$ is said to be {\em an achievable error exponent pair under the expectation constraint} if there exists a sequence of $\{\mathcal{S}_n\}_{n\in\mathbb{N}}$ such that  
\begin{align}
\liminf_{n\to\infty}&\frac{1}{n}\log\frac{1}{\alpha_n}  \ge R_0, \,\liminf_{n\to\infty}\frac{1}{n}\log\frac{1}{\beta_n} \ge  R_1, \quad \mbox{and}\label{eqn:R1} \\
\limsup_{n\to\infty}&\Big( \max_{i \in \{0,1\}} \E_{n,i} [T_{n}]  -n\Big)\le 0.  \label{eqn:exp_con}
\end{align}

Similarly, for $0<\eps<1$, a  pair $(R_0, R_1) \in \mathbb{R}_+^2$ is said to be {\em an $\eps$-achievable error exponent pair under the probabilistic constraint}  if there exists a  sequence of $\{\mathcal{S}_n\}_{n\in\mathbb{N}}$ such that~\eqref{eqn:R1} hold and (instead of\eqref{eqn:exp_con}),
\begin{equation}
\limsup_{n\to\infty} \max_{i \in \{0,1\}} \P_{n,i}(T_{n}>n)< \eps. \label{eqn:prob_con}
\end{equation}
\end{definition}
The condition in \eqref{eqn:exp_con} states that regardless of which hypothesis $i\in  \{0,1\}$ is in effect, the expectation value of the stopping time $\E_{n,i}[T_n]$ should not exceed $n+\gamma$ for any $\gamma>0$ for all $n$ sufficiently large.  In other words, we are allowing some additive slack on $\E_{n,i}[T_n]$.
\begin{definition}[Error Exponent Regions]
Define $\mathcal{A}_\mathrm{E} (\mathcal{N}_{0},\mathcal{N}_{1})\subset\mathbb{R}_+^2$, the {\em error exponent region under the expectation constraint}, to be the closure of the  set of all achievable error exponent pairs under the expectation constraint. 

Similarly, define $\mathcal{A}_\mathrm{P} (\eps|\mathcal{N}_{0},\mathcal{N}_{1})\subset\mathbb{R}_+^2$,  the {\em error exponent region under the $\eps$-probabilistic constraint}, to be the closure of the  set of all $\eps$-achievable error exponent pairs under the probabilistic  constraint.

Define $\mathcal{R}_\mathrm{E}(\mathcal{N}_{0},\mathcal{N}_{1})\subset \mathbb{R}^{2}_{+}$ and $\mathcal{R}_\mathrm{P}(\eps|\mathcal{N}_{0},\mathcal{N}_{1}) \subset\mathbb{R}_+^2$ to be the sets of achievable error exponent pairs using {\em non-adaptive strategies} under the expectation and probabilistic constraints in~\eqref{eqn:exp_con} and~\eqref{eqn:prob_con}, respectively.  
\end{definition}

Throughout the rest of the paper, the quantum channels are fixed, therefore the explicit dependence on the quantum channels is often dropped from the notation for the error exponent regions. 
\subsection{Information Quantities}
Suppose $\rho$ and $\sigma$ are two quantum states over the $d$-dimensional Hilbert space $\mathcal{H}$. The \emph{quantum max-divergence} between two quantum states  $\rho_0$ and $\rho_1$ is defined as
\begin{align}
    D_{\mathrm{max}}(\rho_0\|\rho_1)=\inf\{\gamma:\rho_0\le 2^{\gamma}\rho_1\}.
\end{align}
The {\em quantum relative entropy} between $\rho_0$ and $\rho_1$ is defined as
\begin{align}
    D(\rho_0\|\rho_1)=\Tr[\rho_0(\log\rho_0-\log\rho_1)].
\end{align}
The {\em measured relative entropy} between $\rho$ and $\sigma$ is defined as
\begin{align}\label{eqn:ents}
    D_\mathcal{M}(\rho_0\|\rho_1)=\sup_{m\in\mathcal{P}}D(\P_{\rho_0,m}\|\P_{\rho_1,m}), 
\end{align}
where the supremum is taken over all projector-valued measures $m$. It has been shown in~\cite{Marco2017} that taking the supremum in~\eqref{eqn:ents} with respect to all finite sets $\mathcal{Y}$ and all POVMs with outcomes in $\mathcal{Y}$ gives the same value.

Similarly, the {\em quantum channel relative entropy } between two quantum channels $\mathcal{N}_0$ and $\mathcal{N}_1$ is defined as 
\begin{align}\label{eqn:entc}
    D(\mathcal{N}_0\|\mathcal{N}_1)=\sup_{\rho^{RA}} D\big(\mathcal{N}_0(\rho^{RA}) \big\|\, \mathcal{N}_1(\rho^{RA}) \big),
\end{align}
where the supremum is taken over all bipartite states  $\rho_{RA}$ with an arbitrary ancilla system $R$. Using Schimidt decomposition, one  can show that  the supremum in~\eqref{eqn:entc} can be restricted to pure states $\ket\Psi_{RA}$ with ancilla system $R$ isomorphic to the input system $A$.
The \emph{quantum channel max-divergence} between two quantum channels $\mathcal{N}_0$ and $\mathcal{N}_1$ is defined as
\begin{align}\label{eqn:entcmax}
    D_{\mathrm{max}}(\mathcal{N}_0\|\mathcal{N}_1)=\sup_{\rho^{RA}} D_{\mathrm{max}}\big(\mathcal{N}_0(\rho^{RA}) \big\|\, \mathcal{N}_1(\rho^{RA}) \big),
\end{align}
where the superemum is taken over bipartite states over system $RA$.
The {\em measured relative entropy} between two quantum channels $\mathcal{N}_0$ and $\mathcal{N}_1$ is defined as
\begin{align}\label{eqn:mentcmeasured}
    D_\mathcal{M}(\mathcal{N}_0\|\mathcal{N}_1)=\sup_{\rho^{RA}}D_{\mathcal{M}}(\mathcal{N}_0(\rho^{RA})\|\mathcal{N}_1(\rho^{RA})),
\end{align}
where the supremum is taken over all quantum bipartite states system over $RA$. Similarly to quantum relative entropy between two quantum channels, the supremum is achieved by some pure state $\ket\Psi$ with ancilla system $R$ isomorphic to the input system $A$.

The {\em regularized relative entropy} between two quantum channels is defined as 
\begin{align}
    D^\infty(\mathcal{N}_0\|\mathcal{N}_1))=\lim_{n\to\infty}\frac{1}{n}D(\mathcal{N}^{\otimes n}\|\mathcal{N}_1^{\otimes n}).
\end{align}

\section{Main Results}\label{sec:mainresults}

\subsection{Error Exponents For Adaptive Testing Strategies}

Now we state our main result for the sequential quantum channel discrimination problem.
As the proof of the main results are similar to the results in~\cite{li2021optimal}, we only highlight the main steps of the proof and the differences to the state case in the previous work. Note also that the state case can be a see as a special case of the present setup when both channels have a constant output.

\begin{thm}\label{expectation}
Let $\mathcal{N}_0$ and $\mathcal{N}_1$ be two quantum channels such that $\max_{i=0,1} D_{\mathrm{max}}(\mathcal{N}_i\|\mathcal{N}_{1-i})<\infty$. Then for any $0<\eps<1$,
\begin{align}
\mathcal{A}_\mathrm{P}(\eps)=\mathcal{A}_\mathrm{E}=\left\{(R_0,R_1):\begin{array}{c}
R_0\le D_\mathcal{M}(\mathcal{N}_1\|\mathcal{N}_0)\\
R_1\le D_\mathcal{M}(\mathcal{N}_0\|\mathcal{N}_1)
\end{array}\right\}.
\end{align}
\end{thm}

\begin{proof} We first prove that the corner point of the region $(D_\mathcal{M}(\mathcal{N}_1\|\mathcal{N}_0), D_\mathcal{M}(\mathcal{N}_0\|\mathcal{N}_1))$ is achievable for both types of constraints.  

The adaptive strategy is a variant of the sequential probability ratio test~\cite{Wald1945}. Let $\mathcal{Y}=\{1,2,\ldots,d_{A}d_{B}\}$. From the definition of the measured relative entropy, there exist two input states $\rho_{0}^{*}$ and $\rho_{1}^{*}$ over $RA$ and two PVMs $m^{*}_0=\{m^{*}_{0}(y)\}_{y\in\mathcal{Y}}$ and $m^*_1=\{m^*_{1}(y)\}_{y\in\mathcal{Y}}$ that achieve the supremum in the definitions of $D_\M(\caln_0\|\mathcal{N}_1)$ and $D_\M(\mathcal{N}_1\|\caln_0)$, respectively. 

 We now define the adaptive strategies used in the sequential quantum channel discrimination problem. For $k=1$ and $i\in\{0,1\}$, we set $p_1(\rho_{i})={1}/{2}$ and $\mu(m_{i}^{*}|\rho_{i}^{*})=1$. That is, the experimenter at time $1$ chooses the input state $\rho_{1}\in\{\rho_0^*,\rho_1^*\}$ uniformly at random and then asks the Oracle to process the input quantum state $\rho$ and then applies the measurement corresponding to the input state and obtains the outcome $Y_{1}$.
For $j\ge 2$, we define
\begin{align}
Z_j&=\log{\Tr \big[\mathcal{N}(\rho_j) M_{j}(Y_j) \big]}-\log{\Tr \big[\mathcal{M}(\rho_j) M_{j}(Y_j)\notag \big]}\end{align}
and
$S_{k}=\sum_{j=1}^{k}Z_{j}.$ 
For $k\ge 2$, the input state $\rho_{k}$ and the POVM $M_{k}$ are chosen by the experimenter at time $k$ according to the sign of the accumulated sum of log-likelihoods $S_{k}$ as follows
\begin{align}
\rho_{k}=\begin{cases}
\rho^*_0&\mbox{if}\ S_{k-1}\ge0\\
\rho^*_1&\mbox{otherwise}.
\end{cases}\quad\mbox{and}\quad M_{k}=\begin{cases}
m^*_0&\mbox{if}\ S_{k-1}\ge0\\
m^*_1&\mbox{otherwise}.
\end{cases}\notag
\end{align}
Therefore for $k\ge 1$, the adaptive strategies are defined as follows
\begin{align}
p_{k}(\rho_{0}^{*}, m_0^{*}|\rho_{1}^{k-1},m_{1}^{k-1})&=\begin{cases}\frac{1}{2}&\mbox{if $k=1$}\\1&\mbox{if $k\ge 2$ and $S_{k-1}\ge 0$}\\
0&\mbox{if $k\ge 2$ and $S_{k-1}< 0$},\end{cases} \notag
\end{align}
and
\begin{align} 
p_{k}(\rho_{1}^{*}, m_1^{*}|\rho_{1}^{k-1},m_{1}^{k-1})&=\begin{cases}\frac{1}{2}&\mbox{if $k=1$}\\1&\mbox{if $k\ge 2$ and $S_{k-1}< 0$}\\
0&\mbox{if $k\ge 2$ and $S_{k-1}\ge 0$}.\end{cases}\notag
\end{align} 
 For any fixed $0<\tau<\min\{D_\M(\mathcal{N}_0\|\mathcal{N}_1), D_\M(\mathcal{N}_1\|\mathcal{N}_0)\}$, let 
$
 A_n:=n(D_\M(\mathcal{N}_1\|\mathcal{N}_0)-\tau)$ and $B_n:=n(D_\M(\mathcal{N}_0\|\mathcal{N}_1)-\tau).
$ The decision functions are defined as
 \begin{align}\label{eqn:sprt}
d_{n,k}(Y_{1}^{k},\rho_{1}^{k},M_{1}^{k})=\begin{cases}
0&S_{k}\ge B_n\\
1&S_{k}\le -A_n\\
*&\mbox{otherwise}.
\end{cases}
 \end{align} 
Let $T_n=\inf\{k\ge 1: S_k\not\in(-A_n,B_n)\}$. For any $n\ge 1$, let $\mathcal{S}_{n}=\big(  \mathcal{Y},\{p_{k},d_{n,k}\}_{k=1}^{\infty},T_{n}\big)$ be the sequential adaptive quantum channel discrimination strategies with parameters $A_{n}$ and $B_{n}$.  Using change-of-measure arguments as in the proof of~\cite[Theorems 3.1]{li2021optimal}, we obtain
\begin{align}
\alpha_{n}\le e^{-A_{n}} \quad\mbox{and}\quad\beta_{n}\le e^{-B_{n}}.
\end{align}
Let $\hat{T}_n=\inf\{k\ge 1: S_k\ge B_n\}$. Following similar steps as in the proof of~\cite[Lemma 5.1]{li2021optimal} we can show that 
\begin{itemize}
    \item[(i)] $\E_0[T_n]\le \E_0[\hat{T}_n]$;
    \item[(ii)] $|\E_0[S_{\hat{T}_n}-\hat{T}_n D_\M(\mathcal{N}_0\|\mathcal{N}_1)]|\le C_0$ for some constant $C_0$;
    \item[(iii)] $\E_0[S_{\hat{T}_n}]\le B_n+C_1$ for some constant $C_2$.
\end{itemize}
Then we have that
\begin{align}
    \E_0[T_n]&\le \E_0[\hat{T}_n]\notag\\
    &\le \frac{|\E_0[S_{\hat{T}_n}-\hat{T}_n D_\M(\mathcal{N}_0\|\mathcal{N}_1)]|+\E_0[S_{\hat{T}_n}]}{D_\M(\mathcal{N}_0\|\mathcal{N}_1)}\notag\\
    &\le \frac{n(D_\M(\mathcal{N}_0\|\mathcal{N}_1)-\tau)+C_0+C_1}{D_\M(\mathcal{N}_0\|\mathcal{N}_1)}<n\notag
\end{align}
for sufficiently large $n$. Similarly, we can show that $\E_{i}[T_{n}]<n$ for sufficiently large $n$. This completes the proof of the achievability of the corner point under the expectation constraints.

Note that
\begin{align}
\P_{0}(T_{n}>n)&<\P_0(\hat{T}_n<n)\le \P_0(S_n\le B_n)\notag\\
&\le \frac{\E_0[(S_n-nD_\M(\mathcal{N}_0\|\mathcal{N}_1))^2]}{n^2\tau^2}\\
&\to 0,\label{eqn:prob}
\end{align}
where~\eqref{eqn:prob} follows from the same arguments as in the proof of~~\cite[Equation (136)]{li2021optimal}. This shows that for sufficiently large $n$, $\max_{i=0,1}\P_i(T_n>n)<\eps$. 
Therefore, the sequence of tests $\{\mathcal{S}_{n}\}$ achieves the corner points of regions of error exponents under the probabilistic constraints. 

Now we prove the converse part. Let $\{\alpha_n\}_{n=1}^{\infty}$ and $\{\beta_n\}_{n=1}^{\infty}$ be the type-I and type-II error probabilities of a sequence of sequential tests $\{\mathcal{S}_n\}_{n=1}^{\infty}$ with $\mathcal{S}_n=(\{(\mu_{n,k},\mu_{n,k},d_{n,k})\}_{k=1}^{\infty},T_n)$.  We now sketch the proof of the converse part under the expectation constraints. Using date processing inequality, we have that
\begin{align}\label{eqn:exp:dpi}
\alpha_n\log\frac{\alpha_n}{1-\beta_n}+(1-\alpha_n)\log\frac{1-\alpha_n}{\beta_n}\le \E_{0,n}[S_{T_n}].
\end{align} As $\{S_{n,k}-kD_\M(\mathcal{N}_0\|\mathcal{N}_1)\}_{k=1}^{\infty}$ is a supermartingale, it then follows from the optional stopping theorem that
\begin{align}\label{eqn:exp:ost}
    \E_{0,n}[S_{n,T_n}-T_n D_\M(\mathcal{N}_0\|\mathcal{N}_1)]\le 0.
\end{align}
Combining~\eqref{eqn:exp:dpi} and~\eqref{eqn:exp:ost}, after some elementary algebraic manipulations, we have that
\begin{align}
\limsup_{n\to\infty}\frac{1}{n}\log\frac{1}{\beta_n}\le D_\M(\mathcal{N}_0\|\mathcal{N}_1).\notag
\end{align}
Similar arguments can be applied to the type-I error probability $\alpha_n$. This completes the proof of the converse under the expectation constraints.

Now we sketch the proof of the converse part under the probabilistic constraints. Let $\tau>0$ and let $\lambda_n=n(D_\M(\mathcal{N}_0\|\mathcal{N}_1)+\tau)$. Note that
\begin{align}
 (1-\alpha_n)-e^{\lambda_n}\beta_n&\le \P_{0,n}(S_{n,T_n}\ge \lambda_n)\notag\\
 &\le \P_{0,n}(T_n>n )+\P_{0,n}\Big(\max_{1\le k\le n}S_{n,k}\ge \lambda_n\Big).\notag
\end{align}
Using similar arguments as in~\cite[Equations (155)--(158)]{li2021optimal}, we can show that $\P_{0,n}(\max_{1\le k\le n}S_{n,k}\ge \lambda_n)\le C_3^n$ for some constant $0<C_3<1$. Then after simple algebraic manipulations we have that
\begin{align}
\limsup_{n\to\infty}\frac{1}{n}\log\frac{1}{\beta_n}\le D_\M(\mathcal{N}_0\|\mathcal{N}_1)+\tau.\notag
\end{align}
Simple arguments can be applied to $\alpha_n$. This completes the proof of the converse under the probabilistic constraints. 
\end{proof}
When the experimenter has access to a large quantum memory system, they may each time prepare a state $\rho$ over $R_lA^{\otimes l}$ and ask the Oracle to process $\rho$ using $l$ copies of the underlying channel. Then the experimenter may apply a joint measurement to the output state over $R_lB^{\otimes l}$. In this setting, the problem is equivalent to the binary quantum hypothesis test, 
\begin{align}
H_0^{(l)}: \mathcal{W}^{\otimes l}=\mathcal{N}_0^{\otimes l}\qquad H_1^{(l)}: \mathcal{W}^{\otimes l}=\mathcal{N}_1^{\otimes l}.
\end{align} Under this setup, as was done in Section~\ref{sec:model}, we can define the achievable regions of the error exponent pairs $\mathcal{A}_\mathrm{E}^{(l)}$ and $\mathcal{A}_\mathrm{P}^{(l)}(\eps)$ under the expectation and probabilistic constraints, respectively. Then 
\begin{align}\label{eqn:ultimate}
    \bigcup_{l=1}^{\infty}\mathcal{A}_\mathrm{E}^{(l)}\quad\mbox{and}\quad \bigcup_{l=1}^{\infty}\mathcal{A}_\mathrm{P}^{(l)} 
\end{align}
are the ultimate regions of error exponents pairs that can be achieved by sequentially block adaptive strategies. Our next main results characterize the regions in~\eqref{eqn:ultimate}.
\begin{thm}\label{ultimate:adaptive}
Let $\mathcal{N}_0$ and $\mathcal{N}_1$ be two quantum channels such that $\max_{i=0,1} D_{\mathrm{max}}(\mathcal{N}_i\|\mathcal{N}_{1-i})<\infty$. Then for any $0<\eps<1$, 
\begin{align}\label{eqn:thm:ultimate}
\bigcup_{l=1}^{\infty}\mathcal{A}_\mathrm{E}^{(l)}=\bigcup_{l=1}^{\infty}\mathcal{A}_\mathrm{P}^{(l)}(\eps)=\left\{(R_0,R_1):\begin{array}{c}
R_0 \le D^\infty(\mathcal{N}_1\|\mathcal{N}_0)\\
\, R_1 \le D^\infty(\mathcal{N}_0\|\mathcal{N}_1))
\end{array}\right\}.
\end{align}
\end{thm}
\begin{proof}
Similar to Theorem~\ref{expectation}, we have  
\begin{align}\label{reg:multiple}
\mathcal{A}_\mathrm{E}^{(l)}&=\mathcal{A}_\mathrm{P}^{(l)}(\eps)\notag\\
&=\left\{(R_0,R_1):\begin{array}{c}
R_0\le \displaystyle\frac{1}{l} {D_\mathcal{M}(\mathcal{N}_1^{\otimes l}\|\mathcal{N}_0^{\otimes l})}\vspace{.5em}\\
R_1\le \displaystyle\frac{1}{l} {D_\mathcal{M}(\mathcal{N}_0^{\otimes l}\|\mathcal{N}_1^{\otimes l})}
\end{array}\right\}.
\end{align}

Given $\delta>0$, there exists an integer $K$ and an input state $\rho_{0,K}$ and $\rho_{1,K}$ over the input system $R_{K} A^{\otimes K}$ such that
\begin{align}
D^\infty(\mathcal{N}_0\|\mathcal{N}_1))&\le \frac1K D(\mathcal{N}^{\otimes K}(\rho_{0,K})\|\mathcal{N}_1^{\otimes K}(\rho_{0, K}))+\delta\label{eqn:multiple1}\\
D^\infty(\mathcal{N}_1\|\mathcal{N}_0)&\le \frac1K D(\mathcal{N}_1^{\otimes K}(\rho_{1,K})\|\mathcal{N}^{\otimes K}(\rho_{1,K}))+\delta.\label{eqn:multiple2}
\end{align}

Also, from~\cite{Petz1991}, it follows that
\begin{align}
\lim_{l\to\infty} \frac{1}{l}D_{\mathcal{M}}&(\mathcal{N}_0^{\otimes lK}(\rho_{0,K}^{\otimes l})\|\mathcal{N}_1^{\otimes lK}(\rho_{0, K}^{\otimes l}))\notag\\
&\hspace{0.4cm}=D(\mathcal{N}_0^{\otimes K}(\rho_{0,K})\|\mathcal{N}_1^{\otimes K}(\rho_{0, K}))\label{eqn:m1}
\end{align}
and
\begin{align}\lim_{l\to\infty} \frac{1}{l}D_{\mathcal{M}}&(\cN_1^{\otimes lK}(\rho_{1,K}^{\otimes l})\|\mathcal{N}_0^{\otimes lK}(\rho_{1, K}^{\otimes l}))\notag\\
&=D(\cN_1^{\otimes K}(\rho_{1,K})\|\mathcal{N}_0^{\otimes K}(\rho_{1,K}))\label{eqn:m2}.
\end{align}
Then from~\eqref{eqn:multiple1},~\eqref{eqn:multiple2},~\eqref{eqn:m1}, and~\eqref{eqn:m2} we obtain
{\small\begin{align}
&\left\{(R_0,R_1):\begin{array}{c}
R_0\le D^\infty(\cN_1\|\mathcal{N}_0)-\delta\vspace{.5em}\\
R_1\le D^\infty(\mathcal{N}_0\|\cN_1)-\delta
\end{array}\right\}\notag\\
&\subset \left\{(R_0,R_1):\begin{array}{c}
R_0\le \frac{1}{K}D(\cN_1^{\otimes K}(\rho_{1,K})\|\mathcal{N}_0^{\otimes K}(\rho_{1, K}))\vspace{.5em}\\
R_1\le \frac{1}{K}D(\mathcal{N}_0^{\otimes K}(\rho_{0,K})\|\cN_1^{\otimes K}(\rho_{0, K}))
\end{array}\right\}\\
&\subset \bigcup_{l=1}^{\infty}\left\{(R_0,R_1):\begin{array}{c}
R_0\le \frac{1}{lK}D_{\mathcal{M}}(\cN_1^{\otimes lK}(\rho_{1,K}^{\otimes l})\|\mathcal{N}_0^{\otimes lK}(\rho_{1, K}^{\otimes l}))\\
R_1\le \frac{1}{lK}D_{\mathcal{M}}(\mathcal{N}_0^{\otimes lK}(\rho_{0,K}^{\otimes l})\|\cN_1^{\otimes lK}(\rho_{0, K}^{\otimes l}))
\end{array}\right\}\\
&\subset\bigcup_{l=1}^{\infty}\left\{(R_0,R_1):\begin{array}{c}
R_0\le \frac{1}{lK}D_{\mathcal{M}}(\cN_1^{\otimes lK}\|\mathcal{N}_0^{\otimes lK})\\
R_1\le \frac{1}{lK}D_{\mathcal{M}}(\mathcal{N}_0^{\otimes lK}\|\cN_1^{\otimes lK})
\end{array}\right\}.
\end{align}}
In combination with~(\ref{reg:multiple}), we have that
\begin{align}\label{eqn:limiting}
&\left\{(R_0,R_1):\begin{array}{c}
R_0\le D^\infty(\cN_1\|\mathcal{N}_0)-\delta\vspace{.5em}\\
R_1\le D^\infty(\mathcal{N}_0\|\cN_1)-\delta
\end{array}\right\}\notag\\
&\hspace{3cm}\subset\bigcup_{l=1}^{\infty}\mathcal{A}_{\mathrm{E}}^{(l)}=\bigcup_{l=1}^{\infty}\mathcal{A}_{\mathrm{P}}^{(l)}(\eps).
\end{align}
From the data-processing inequality~\cite[Chapter 11]{Wilde2013} for the  quantum relative entropy, we have that
\begin{align}\label{eqn:dpi}
D_{\mathcal{M}}&(\cN_1^{\otimes lK}\|\mathcal{N}_0^{\otimes lK})\le D(\cN_1^{\otimes lK}\|\mathcal{N}_0^{\otimes lK}),\\
\mbox{and}\quad D_{\cN_1}&(\mathcal{N}_0^{\otimes lK}\|\cN_1^{\otimes lK})\le D(\mathcal{N}_0^{\otimes lK}\|\cN_1^{\otimes lK}).
\end{align}
Combining~\eqref{eqn:limiting} with~\eqref{eqn:dpi}, we have~\eqref{eqn:thm:ultimate} as desired.
\end{proof}

\subsection{Error Exponent Regions with Non-Adaptive Testing Strategies} \label{sec:non-adapt}
In this section we state our results for $\mathcal{R}_{\mathrm{E}}$ and $\mathcal{R}_{\mathrm{P}}(\eps)$, the regions of error exponent pairs when {\em non-adaptive} tests are permitted. For any subset $A$ of the plane $\mathbb{R}^{2}$, let $\overline{\mathrm{Conv}(A)}$ be the closure of the convex hull of $A$.  Given an input state $\rho$, a quantum channel $\mathcal{W}$ and a POVM $m\in\mathcal{Y}$, we define
\begin{align}
P_{\mathcal{W},\rho,m}(y)=\Tr[\mathcal{W}(\rho)m_y].
\end{align} In other words, $P_{\mathcal{W},\rho,m}$ is the probability mass function of the outcome obtained when the POVM $m$ is applied to the output of the quantum channel $\W$ with the input state $\rho$. The following  theorem characterizes   $\mathcal{R}_\mathrm{E}$ and $\mathcal{R}_\mathrm{P}(\eps)$.

\begin{thm}\label{non-adaptive}
Let $\mathcal{N}_0$ and $\mathcal{N}_1$ be two quantum channels such that $\max_{i=0,1} D_{\mathrm{max}}(\mathcal{N}_i\|\mathcal{N}_{1-i})<\infty$. Then for any $0<\eps<1$, 
\begin{align}
 \mathcal{R}_\mathrm{E}=\mathcal{R}_{\mathrm{P}}(\eps)=\overline{\mathrm{Conv}(\mathcal{C})},
\end{align}
where 
\begin{align}
\hspace{-.1cm}\mathcal{C}=
\bigcup_{\rho^{RA},\mathcal{Y},m }\left\{(R_0,R_1):\begin{array}{c}
R_0\le D(P_{\cN_1,\rho^{RA},m}\|P_{\mathcal{N}_0,\rho^{RA}, m})\\
R_1\le D(P_{\mathcal{N}_0,\rho^{RA},m}\|P_{\cN_1,\rho^{RA}, m})
\end{array}
\right\},\notag
\end{align}
and $\rho^{RA}$ runs over all states on system $RA$, $\mathcal{Y}$ runs over all finite sets and $m$ runs over $\mathcal{M}_{\mathcal{Y}}$, the set of POVMs with support $\mathcal{Y}$.
\end{thm}


\section{Results for general discrimination strategy}
The most general measurements to extract information from a quantum system is a {\em quantum instrument} which, besides providing a measurement outcome, leaves us with a residual state that can be further processed. Therefore, to sequentially discriminate two quantum channels, instead of using a POVM and discarding the residual quantum state, the experimenter may apply a quantum instrument and make use of that residual state during a future measurement. Formally a quantum instrument is defined as follows.   
\begin{definition}
A quantum instrument $\mathcal{I}=\{\mathcal{E}_{y}(\cdot):y\in\mathcal{Y}\}$ is a set of completely positive trace-nonincreasing maps such that $\sum_{y\in\mathcal{Y}}\mathcal{I}_{y}(\cdot)$ is trace-preserving.
\end{definition}
Now we assume that the alphabet $\mathcal{Y}$ is a finite set and assume there is enough quantum memory to store the residual quantum states during the discrimination procedure. We now describe the procedure to sequentially discriminate two quantum channels using quantum instruments and quantum memory. The outcome of each quantum instrument $\mathcal{I}_k=\{\mathcal{E}_y:y\in\mathcal{Y}\}$ is in $\mathcal{Y}$. Each time the quantum instrument measures the output state and output the outcome and the next input state simultaneously. Assume the underlying quantum channel is $\mathcal{N}_{\nu}$. The output state at time $k$ is denoted by $\sigma_{\nu,R_kB_k}$.

At time $1$, the experimenter chooses the input state $\rho_{R_1A_1}$ and the instrument $\mathcal{I}_{1}$. After obtaining the output state $\sigma_{\nu,R_1B_1}$, the experimenter applies the quantum instrument $\mathcal{I}_{1}$ to  $\sigma_{R_1B_1}$ and obtains outcome $y_{1}$ and the next input state $\rho_{R_2A_2}$. The state $\rho_{R_2A_2}$ is stored at some quantum register for later use. Based on $(\rho_{R_1A_1},\mathcal{I}_1,y_1)$, the experimenter applies the decision function $d_1$: if $d_1=i$ for $i\in\{0,1\}$, the experimenter stops the experiment and makes the decision that $\mathcal{N}_i$ is the underlying channels; otherwise the experimenter continues the experiment. For $k\ge 2$, the experimenter prepares the quantum instrument $\mathcal{I}_k$ conditioning on $\rho_{R_1A_1}$, previous quantum instruments $\mathcal{I}_1^{k-1}$ and previous outcomes $y_1^{k-1}$. Then the experimenter passes the quantum state $\rho_{R_kA_k}$ (generated by quantum instrument $\mathcal{I}_{k-1}$) through the underlying quantum channel. After receiving the output state $\sigma_{\nu,R_kB_k}$, the experimenter applies the quantum instrument $\mathcal{I}_k$ to it and obtains the outcome $y_k$ and the new input state $\rho_{R_{k+1}A_{k+1}}$, which is stored in some quantum register for later use. Then the experimenter use the decision function $d_{k}(y_{1}^{k},\rho_{1},\mathcal{I}_{1}^{k})\in\{0,1,*\}$ to decide whether to continue or stop the experiment: if $d_{k}=*$, the experimenter decides to continue the procedure; otherwise, the experimenter stops the procedure and makes the decision that $\mathcal{N}_{d_{k}}$ is the underlying quantum channel. 

The key difference between the testing procedure and the one used in Section~\ref{sec:model} is that the output state after the measurement is stored in some quantum register for later use. Similarly, we can define the region of error exponents under the expectation and probabilistic constraints $\tilde{\mathcal{A}}_{\mathrm{E}}$ and $\tilde{\mathcal{A}}_{\mathrm{P}}(\eps)$. Clearly, this strategy is more general than the adaptive strategy using POVMs. Therefore Theorems~\ref{expectation} and~\ref{ultimate:adaptive} can serve as achievability results in this setting which leads directly to the following result. 
\begin{pr}
Let $\mathcal{N}_0$ and $\mathcal{N}_1$ be two quantum channels such that $\max_{i=0,1} D_{\mathrm{max}}(\mathcal{N}_i\|\mathcal{N}_{1-i})<\infty$. It holds that
\begin{align}
   \left\{(R_0,R_1):\begin{array}{c}
R_0 \le D^\infty(\mathcal{N}_1\|\mathcal{N}_0)\\
 R_1 \le D^\infty(\mathcal{N}_0\|\mathcal{N}_1))
 \end{array}
 \right\}\subset \tilde{\mathcal{A}}_{\mathrm{E}}
\end{align}
and
\begin{align}
   \left\{(R_0,R_1):\begin{array}{c}
R_0 \le D^\infty(\mathcal{N}_1\|\mathcal{N}_0)\\
 R_1 \le D^\infty(\mathcal{N}_0\|\mathcal{N}_1))
 \end{array}
 \right\}\subset \tilde{\mathcal{A}}_{\mathrm{P}}(\eps).
\end{align}
\end{pr} 
The difficulty in proving a converse for the described setting is that the accumulated likelihood 
\begin{align}
S_{l}=\sum_{k=1}^{l}\log\frac{\Tr[\mathcal{E}_{Y_{k}}(\sigma_{0,R_kB_k})]}{\Tr[\mathcal{E}_{Y_{k}}(\sigma_{1,R_kB_k})]}
\end{align}
 is not regular; that is, the absolute value of  $|S_{l}|$ may be close to $0$ if the instruments used only reveal very little information about the underlying channel but $|S_{l}|$ may be large if the instruments used at time $k$ extract much information.  This irregularity of $\{S_{l}\}$ makes the arguments in the proof of Theorem~\ref{expectation} inapplicable. However, we conjecture that the regions are not larger than those achievable with POVMs, i.e.,
 \begin{align}\label{eqn:ourconj}
     \left\{(R_0,R_1):\begin{array}{c}
R_0  \le D^\infty(\mathcal{N}_1\|\mathcal{N}_0)\\
 R_1  \le D^\infty(\mathcal{N}_0\|\mathcal{N}_1)
 \end{array}
 \right\}=\tilde{\mathcal{A}}_{\mathrm{E}} = \tilde{\mathcal{A}}_{\mathrm{P}}(\eps), 
 \end{align}
 implying that the optimal region is determined by the regularized channel relative entropy, similar to the quantum channel Stein's lemma. Actually we can show that this conjecture is implied by the strong converse for quantum channel discrimination problem. 
 While we can not prove this conjecture we can still give a converse bound that we conjecture to coincide with the desired region. 
 To that end we define the sandwiched quantum channel R\'enyi divergence $\tilde D_\alpha(\cN_0\|\cN_1)$ and from it
 \begin{align}
     \tilde D^\infty_\alpha(\cN_0\|\cN_1) &= \lim_{n\rightarrow\infty}\frac1n \tilde D_\alpha(\cN_0^{\otimes n}\|\cN_1^{\otimes n}).
 \end{align}
 \begin{pr}\label{pr:alpha-converse}
Let $\mathcal{N}_0$ and $\mathcal{N}_1$ be two quantum channels such that $\max_{i=0,1} D_{\mathrm{max}}(\mathcal{N}_i\|\mathcal{N}_{1-i})<\infty$. Let
 \begin{align}
 \widehat{\mathcal{R}}=\left\{(R_0,R_1):\begin{array}{c}
 R_0 \le   \lim_{\alpha\rightarrow 1} \tilde D^\infty_\alpha(\cN_1\|\cN_0)\\
 \, R_1 \le  \lim_{\alpha\rightarrow 1} \tilde D^\infty_\alpha(\cN_0\|\cN_1)
 \end{array}\right\}.
 \end{align} Then for any $0<\eps<1$, 
 \begin{align}
 \tilde{\mathcal{A}}_{\mathrm{E}}\subset\widehat{\mathcal{R}}  \quad\mbox{and}\quad \tilde{\mathcal{A}}_{\mathrm{P}}(\eps)\subset\widehat{\mathcal{R}}.
 \end{align}
 \end{pr}
 \begin{remark}
If
 \begin{align}\label{eqn:conj}
      \lim_{\alpha\rightarrow 1} \tilde D^\infty_\alpha(\cN_0\|\cN_1){=} D^\infty(\cN_0\|\cN_1),
 \end{align}
then together with Proposition~\ref{pr:alpha-converse}, we can show equation~\eqref{eqn:ourconj}. However, we do not know whether~\eqref{eqn:conj} holds and if it holds, it would imply a strong converse for the fixed-length quantum channel discrimination problem, which is an important open problem in quantum information theory. For more details, see also~\cite{Fawzi2021}.
 \end{remark}

\begin{proof}
We prove Proposition~\ref{pr:alpha-converse} with expectation constrained through an approach similar to that used in the proof of the converse for general adaptive strategies for sequential hypothesis testing for states in~\cite{quantumSHT} and with probability constrained inspired by a recent result in~\cite{fanizza2022qusum}. We will split the proof into two parts, first proving a general lower bound for $\P_{i}(T_n>k)$ in terms of error probabilities and then proving the converse under different types of constraints.
Let $\{\mathcal{S}_n\}_{n=1}^{\infty}$ be a sequence of a  sequential channel discrimination strategy with  error probabilities  $\{\alpha_n\}_{n=1}^{\infty}$ and $\{\beta_n\}_{n=1}^{\infty}$. For $i=0,1$ and any $j\ge1$, let $C_{i,j}$ be the event that $Y_1^{j-1}=2$ and $Y_j=i$. Then we have that
\begin{align}
    \P_{i,n}(T_n>k)=1-\P_{i,n}(T_n\le k)=1-\sum_{j=1}^k\big(\P_{i,n}(C_{0,j})+\P_{i,n}(C_{1,j})\big).
\end{align}
Consider the adaptive strategies $\mathcal{S}_n$ applied to $\mathcal{N}_{\nu}^{\otimes k}$: if $d_j=0$ for some $j\le k$, then the experimenter makes the decision that $\mathcal{N}_0$ is the underlying channel; otherwise at time $k$ the experimenter makes the decision that $\mathcal{N}_1$ is the underlying channel. Let $\tilde{\alpha}_{n,k}$ and $\tilde{\beta}_{n,k}$ be the type-I and type-II error probabilities of this testing strategy. Then we have that
\begin{align}\label{property:fixed}
\tilde{\alpha}_{n,k}&=\P_{0,n}(T_n>k)+\sum_{j=1}^k \P_{0,n}(C_{1,j})\\
\tilde{\beta}_{n,k}&=\sum_{j=1}^k \P_{1,n}(C_{0,j}),
\end{align}
and
\begin{align}
 \sum_{j=1}^k \P_{0,n}(C_{1,j})\le \alpha_n,\quad \tilde{\beta}_{n,k}\le \beta_n\quad\mbox{and}\quad\lim_{k\to\infty}\tilde{\beta}_{n,k}=\beta_n.
\end{align}
Therefore, we have that
\begin{align}\label{eqn:lb:excess}
     \P_{0,n}(T_n>k)=1-\P_{0,n}(T_n\le k)=\tilde{\alpha}_{n,k}-\sum_{j=1}^n\P_0(C_{1,j})\ge \tilde{\alpha}_{n,k}-\alpha_n.
\end{align}
Now we derive a lower bound for $\tilde{\alpha}_{n,k}$ in terms of $\beta_n$. We will use two main ingredients. The first one is the following inequality proven in~\cite[Equation (147)]{Wilde2020} that for any $\alpha>1$,
\begin{align}
    -\frac1k \log(1-\tilde{\alpha}_{n,k}) \geq \frac{\alpha-1}{\alpha}\bigg(-\frac1k\log\tilde{\beta}_{n,k} - \tilde D^A_\alpha(\cN_0\|\cN_1)\bigg),
\end{align}
where $\tilde D^A_\alpha(\cN_0\|\cN_1)$ is the amortized R\'enyi channel relative entropy. The second one is that $\tilde D^A_\alpha(\cN_0\|\cN_1)=\tilde D^\infty_\alpha(\cN_0\|\cN_1)$, which was recently shown in~\cite{Fawzi2021}. Therefore for any $\alpha>1$, we have that
\begin{align}
    -\frac1k \log(1-\tilde{\alpha}_{n,k}) \geq \frac{\alpha-1}{\alpha}\bigg(-\frac1k\log\tilde{\beta}_{n,k} - \tilde D^\infty_\alpha(\cN_0\|\cN_1)\bigg). 
\end{align}
Simple calculations show that
\begin{align}
\tilde{\alpha}_{n,k}&\ge 1-e^{-\frac{\alpha-1}{\alpha}(-\log\tilde{\beta}_{n,k}-k\tilde D^\infty_\alpha(\cN_0\|\cN_1))}\\
&\ge 1-e^{-\frac{\alpha-1}{\alpha}(-\log \beta_{n}-k\tilde D^\infty_\alpha(\cN_0\|\cN_1))}.\label{eqn:converse}
\end{align}

Now we prove the converse under the expectation constraint. Let $\{\mathcal{S}_n\}_{n=1}^{\infty}$ be a sequence of sequential channel discrimination strategies satisfying the expectation constraints and that the  error probabilities  $\{\alpha_n\}_{n=1}^{\infty}$ and $\{\beta_n\}_{n=1}^{\infty}$ are such that $\max\{\alpha_n,\beta_n\}\to 0$ as $n\to\infty$. For $0<\tau<1$, let 
\begin{align}
   k\triangleq k_n=-\frac{\log\beta_n}{\tilde{D}_\alpha^{\infty}(\caln_0\|\caln_1)+\tau}.
\end{align} Then we have that
\begin{align}
    n&\ge \E_{0,n}[T_n]\\
    &\ge \left(\P_{0,n}\bigg(T_n>\frac{-\log\beta_n}{\tilde{D}_\alpha^{\infty}(\caln_0\|\caln_1)+\tau}\bigg)\right)\frac{-\log\beta_n}{\tilde{D}_\alpha^{\infty}(\caln_0\|\caln_1)+\tau}\label{eqn:exp:markov}\\
&\ge \left(1-e^{-\frac{\alpha-1}{\alpha}(-\log \beta_{n}-k_n\tilde D^\infty_\alpha(\cN_0\|\cN_1))}-\alpha_n\right)\frac{-\log\beta_n}{\tilde{D}_\alpha(\caln_0\|\caln_1)+\tau_1}\label{eqn:exp:con:lower}\\
&=\left(1-\beta_n^{-\frac{\alpha-1}{\alpha}(-\log \beta_{n}-k_n\tilde D^\infty_\alpha(\cN_0\|\cN_1))}-\alpha_n\right)\frac{-\log\beta_n}{\tilde{D}_\alpha(\caln_0\|\caln_1)+\tau_1}\\
    &\ge \left(1-\beta_n^{\frac{(\alpha-1)\tau}{\alpha(\tilde{D}_\alpha^{\infty}(\caln_0\|\caln_1)+\tau)}}-\alpha_n\right)\frac{-\log\beta_n}{\tilde{D}_\alpha(\caln_0\|\caln_1)+\tau_1},
\end{align}
where~\eqref{eqn:exp:markov} follows from Markov inequality and \eqref{eqn:exp:con:lower} follows from~\eqref{eqn:converse}.
Then for any $\alpha>1$, it follows that
\begin{align}
\limsup_{n\to\infty}\frac1n\log\frac{1}{\beta_n}\le \tilde{D}_\alpha^{\infty}(\caln_0\|\caln_1)+\tau.
\end{align}
Due to the arbitrariness of $\tau$, it follows that
\begin{align}
\limsup_{n\to\infty}\frac1n\log\frac{1}{\beta_n}\le \tilde{D}_\alpha^{\infty}(\caln_0\|\caln_1).
\end{align}
As the above equation holds for any $\alpha>1$, it holds that
\begin{align}
    \limsup_{n\to\infty}\frac{1}{n}\log\frac{1}{\beta_n}&\le  \lim_{\alpha\rightarrow 1} \tilde D^\infty_\alpha(\cN_0\|\cN_1).
\end{align}
Similar arguments show that
\begin{align}
    \limsup_{n\to\infty}\frac{1}{n}\log\frac{1}{\alpha_n}&\le  \lim_{\alpha\rightarrow 1} \tilde D^\infty_\alpha(\cN_1\|\cN_0),
\end{align}
which completes the proof of the converse part under the expectation constraints.

Now we prove the converse under the probabilistic constraint. Let $\{\mathcal{S}_n\}_{n=1}^{\infty}$ be a sequence of sequential channel discrimination strategies satisfying the probabilistic constraints and that the  error probabilities  $\{\alpha_n\}_{n=1}^{\infty}$ and $\{\beta_n\}_{n=1}^{\infty}$ are such that $\max\{\alpha_n,\beta_n\}\to 0$ as $n\to\infty$. Letting $k=n$, we have that
\begin{align}
    \eps&\ge \P_{0,n}(T_n\ge n)\\
    &\ge  1-e^{-\frac{\alpha-1}{\alpha}(-\log \beta_{n}-n\tilde D^\infty_\alpha(\cN_0\|\cN_1))}-\alpha_n,\label{eq:prob:conv}
\end{align}
where~\eqref{eq:prob:conv} follows from~\eqref{eqn:converse}.
After simple manipulations, we obtain
\begin{align}
    \limsup_{n\to\infty}\frac{1}{n}\log\frac{1}{\beta_n}&\le \limsup_{n\to\infty}-\frac{\alpha}{n(1-\alpha)}\log(1-\eps-\alpha_n)+\tilde{D}_\alpha^{\infty}(\caln_0\|\caln_1)\\
    &=\tilde{D}_\alpha^{\infty}(\caln_0\|\caln_1).
\end{align}
As the above equation holds for any $\alpha>1$, it holds that
\begin{align}
    \limsup_{n\to\infty}\frac{1}{n}\log\frac{1}{\beta_n}&\le  \lim_{\alpha\rightarrow 1} \tilde D^\infty_\alpha(\cN_0\|\cN_1).
\end{align}
Similar arguments show that
\begin{align}
    \limsup_{n\to\infty}\frac{1}{n}\log\frac{1}{\alpha_n}&\le  \lim_{\alpha\rightarrow 1} \tilde D^\infty_\alpha(\cN_1\|\cN_0),
\end{align}
which completes the proof of the converse part under the probabilistic constraints.

\end{proof}

\section{Conclusion  and Discussion}
In this paper, we consider the sequential quantum channel discrimination problem and characterize the error exponent pairs under different types of constraints on sample size and different types of strategies used in the quantum channel discrimination problem. Most notably, using adaptive strategies the exponent pairs are given by the regularized channel relative entropy. 

However, the most general measurements to extract information from
a quantum system is a {\em quantum instrument}. Therefore, to sequentially discriminate two quantum channels, instead using a POVM each time to measure the output system and discard the residual quantum state, the experimenter may apply a quantum instrument and leave the residual state for future use. We conjecture that the achievable region of error exponents pairs obtained by adaptively choosing quantum instruments is the same as the region obtained in Theorem~\ref{ultimate:adaptive}. However, the stochastic process obtained by adaptively applying quantum instruments to the underlying quantum system is not a {\em martingale} anymore, the tools developed for the other settings fail and new ideas are needed. 
We use similar approaches as used in~\cite{quantumSHT,fanizza2022qusum} to prove converse bounds in the form of regularized channel sandwiched \Renyi entropies. Whether these converse bounds coincide with the achievability results provided is the biggest open problem left in sequential quantum channel discrimination and is equivalent to proving a strong converse for the quantum channel Stein's Lemma.


\section*{Acknowledgments} 
YL is supported a National Research Foundation Fellowship  (R-263-000-D02-281). MT and CH are supported by the NRF, Prime
Minister’s Office, Singapore and the Ministry of Education, Singapore under the Research Centres of Excellence programme. CH has received funding from the European Union's Horizon 2020 research and innovation programme under the Marie Sklodowska-Curie Grant Agreement No. H2020-MSCA-IF-2020-101025848. MT and YL are also supported by NUS startup grants (R-263-000-E32-133 and R-263-000-E32-731).

\bibliographystyle{IEEEtran}
\bibliography{reference_2}

\begin{thebibliography}{10}
\providecommand{\url}[1]{#1}
\csname url@samestyle\endcsname
\providecommand{\newblock}{\relax}
\providecommand{\bibinfo}[2]{#2}
\providecommand{\BIBentrySTDinterwordspacing}{\spaceskip=0pt\relax}
\providecommand{\BIBentryALTinterwordstretchfactor}{4}
\providecommand{\BIBentryALTinterwordspacing}{\spaceskip=\fontdimen2\font plus
\BIBentryALTinterwordstretchfactor\fontdimen3\font minus
  \fontdimen4\font\relax}
\providecommand{\BIBforeignlanguage}[2]{{%
\expandafter\ifx\csname l@#1\endcsname\relax
\typeout{** WARNING: IEEEtran.bst: No hyphenation pattern has been}%
\typeout{** loaded for the language `#1'. Using the pattern for}%
\typeout{** the default language instead.}%
\else
\language=\csname l@#1\endcsname
\fi
#2}}
\providecommand{\BIBdecl}{\relax}
\BIBdecl

\bibitem{audenaert07-3}
K.~M.~R. Audenaert, M.~Nussbaum, A.~Szko{\l}a, and F.~Verstraete, ``{Asymptotic
  error rates in quantum hypothesis testing},'' \emph{Communications in
  Mathematical Physics}, vol. 279, no.~1, pp. 251--283, 2008.

\bibitem{Petz1991}
F.~Hiai and D.~Petz, ``The proper formula for relative entropy and its
  asymptotics in quantum probability,'' \emph{Communications in Mathematical
  Physics}, vol. 143, no.~1, pp. 99--114, 1991.

\bibitem{Nagaoka2000}
T.~{Ogawa} and H.~{Nagaoka}, ``Strong converse and {Stein's} lemma in quantum
  hypothesis testing,'' \emph{IEEE Transactions on Information Theory},
  vol.~46, no.~7, pp. 2428--2433, 2000.

\bibitem{Hayashi2004}
T.~{Ogawa} and M.~{Hayashi}, ``On error exponents in quantum hypothesis
  testing,'' \emph{IEEE Transactions on Information Theory}, vol.~50, no.~6,
  pp. 1368--1372, 2004.

\bibitem{Nagaoka2006}
\BIBentryALTinterwordspacing
H.~Nagaoka, ``The converse part of the theorem for quantum {Hoeffding} bound,''
  \emph{arXiv: Quantum Physics}, 2006. [Online]. Available:
  \url{https://arxiv.org/abs/quant-ph/0611289}
\BIBentrySTDinterwordspacing

\bibitem{Hayashi2007}
M.~Hayashi, ``Error exponent in asymmetric quantum hypothesis testing and its
  application to classical-quantum channel coding,'' \emph{Physical Review A},
  vol.~76, no.~6, p. 062301, 2007.

\bibitem{Audenaert2007}
K.~M.~R. Audenaert, J.~Calsamiglia, R.~Mu\~noz Tapia, E.~Bagan, L.~Masanes,
  A.~Acin, and F.~Verstraete, ``Discriminating states: The quantum {Chernoff}
  bound,'' \emph{Physical Review Letters}, vol.~98, no.~16, p. 160501, 2007.

\bibitem{Nussbaum2009}
M.~Nussbaum and A.~Szko{\l}a, ``{The Chernoff lower bound for symmetric quantum
  hypothesis testing},'' \emph{The Annals of Statistics}, vol.~37, no.~2, pp.
  1040--1057, 2009.

\bibitem{berta2021composite}
M.~Berta, F.~G. Brandao, and C.~Hirche, ``On composite quantum hypothesis
  testing,'' \emph{Communications in Mathematical Physics}, vol. 385, no.~1,
  pp. 55--77, 2021.

\bibitem{Wald1945}
A.~Wald, ``{Sequential tests of statistical hypotheses},'' \emph{The Annals of
  Mathematical Statistics}, vol.~16, no.~2, pp. 117--186, 1945.

\bibitem{quantumSHT}
\BIBentryALTinterwordspacing
E.~Mart\'{\i}nez~Vargas, C.~Hirche, G.~Sent\'{\i}s, M.~Skotiniotis, M.~Carrizo,
  R.~Mu\~noz Tapia, and J.~Calsamiglia, ``Quantum sequential hypothesis
  testing,'' \emph{Phys. Rev. Lett.}, vol. 126, p. 180502, May 2021. [Online].
  Available: \url{https://link.aps.org/doi/10.1103/PhysRevLett.126.180502}
\BIBentrySTDinterwordspacing

\bibitem{li2021optimal}
\BIBentryALTinterwordspacing
Y.~Li, V.~Y.~F. Tan, and M.~Tomamichel, ``Optimal adaptive strategies for
  sequential quantum hypothesis testing,'' 2021. [Online]. Available:
  \url{https://arxiv.org/abs/2104.14706}
\BIBentrySTDinterwordspacing

\bibitem{Wilde2020}
\BIBentryALTinterwordspacing
M.~M. Wilde, M.~Berta, C.~Hirche, and E.~Kaur, ``Amortized channel divergence
  for asymptotic quantum channel discrimination,'' \emph{Letters in
  Mathematical Physics}, vol. 110, no.~8, pp. 2277--2336, 2020. [Online].
  Available: \url{https://doi.org/10.1007/s11005-020-01297-7}
\BIBentrySTDinterwordspacing

\bibitem{berta2019stein}
M.~Berta, C.~Hirche, E.~Kaur, and M.~M. Wilde, ``Stein’s lemma for
  classical-quantum channels,'' in \emph{2019 IEEE International Symposium on
  Information Theory (ISIT)}.\hskip 1em plus 0.5em minus 0.4em\relax IEEE,
  2019, pp. 2564--2568.

\bibitem{wang2019resource}
X.~Wang and M.~M. Wilde, ``Resource theory of asymmetric distinguishability for
  quantum channels,'' \emph{Physical Review Research}, vol.~1, no.~3, p.
  033169, 2019.

\bibitem{fang2020chain}
K.~Fang, O.~Fawzi, R.~Renner, and D.~Sutter, ``Chain rule for the quantum
  relative entropy,'' \emph{Physical review letters}, vol. 124, no.~10, p.
  100501, 2020.

\bibitem{tomamichel12}
M.~Tomamichel and M.~Hayashi, ``{A hierarchy of information quantities for
  finite block length analysis of quantum tasks},'' \emph{IEEE Transactions on
  Information Theory}, vol.~59, no.~11, pp. 7693--7710, 2013.

\bibitem{li12}
K.~Li, ``{Second-order asymptotics for quantum hypothesis testing},''
  \emph{Annals of Statistics}, vol.~42, no.~1, pp. 171--189, 2014.

\bibitem{chubb17}
C.~T. Chubb, V.~Y.~F. Tan, and M.~Tomamichel, ``{Moderate deviation analysis
  for classical communication over quantum channels},'' \emph{Communications in
  Mathematical Physics}, vol. 355, no.~3, pp. 1283--1315, 2017.

\bibitem{cheng17}
H.-C. Cheng and M.-H. Hsieh, ``{Moderate deviation analysis for
  classical-quantum channels and quantum hypothesis testing},'' \emph{IEEE
  Transactions on Information Theory}, vol.~64, no.~2, pp. 1385--1403, 2018.

\bibitem{Hayashi2009}
M.~{Hayashi}, ``Discrimination of two channels by adaptive methods and its
  application to quantum system,'' \emph{IEEE Transactions on Information
  Theory}, vol.~55, no.~8, pp. 3807--3820, 2009.

\bibitem{Wilde2016}
\BIBentryALTinterwordspacing
T.~Cooney, M.~Mosonyi, and M.~M. Wilde, ``Strong converse exponents for a
  quantum channel discrimination problem and quantum-feedback-assisted
  communication,'' \emph{Communications in Mathematical Physics}, vol. 344,
  no.~3, pp. 797--829, 2016. [Online]. Available:
  \url{https://doi.org/10.1007/s00220-016-2645-4}
\BIBentrySTDinterwordspacing

\bibitem{Fawzi2021}
H.~Fawzi and O.~Fawzi, ``Defining quantum divergences via convex
  optimization,'' \emph{Quantum}, vol.~5, p. 387, 2021.

\bibitem{WaldWolf}
A.~Wald and J.~Wolfowitz, ``Optimum character of the sequential probability
  ratio test,'' \emph{The Annals of Mathematical Statistics}, vol.~19, no.~3,
  pp. 326--339, 1948.

\bibitem{Hoeffding1965}
W.~Hoeffding, ``{Asymptotically optimal tests for multinomial distributions},''
  \emph{The Annals of Mathematical Statistics}, vol.~36, no.~2, pp. 369--401,
  1965.

\bibitem{Blahut}
R.~Blahut, ``Hypothesis testing and information theory,'' \emph{IEEE
  Transactions on Information Theory}, vol.~20, no.~7, pp. 405--417, 1974.

\bibitem{slussarenko17}
\BIBentryALTinterwordspacing
S.~Slussarenko, M.~M. Weston, J.-G. Li, N.~Campbell, H.~M. Wiseman, and G.~J.
  Pryde, ``Quantum state discrimination using the minimum average number of
  copies,'' \emph{Phys. Rev. Lett.}, vol. 118, p. 030502, Jan 2017. [Online].
  Available: \url{https://link.aps.org/doi/10.1103/PhysRevLett.118.030502}
\BIBentrySTDinterwordspacing

\bibitem{HayashiQIT}
\BIBentryALTinterwordspacing
M.~Hayashi, \emph{Quantum Information Theory: Mathematical Foundation}.\hskip
  1em plus 0.5em minus 0.4em\relax Berlin, Heidelberg: Springer Berlin
  Heidelberg, 2017. [Online]. Available:
  \url{https://doi.org/10.1007/978-3-662-49725-8}
\BIBentrySTDinterwordspacing

\bibitem{anusha}
A.~{Lalitha} and T.~{Javidi}, ``Reliability of sequential hypothesis testing
  can be achieved by an almost-fixed-length test,'' in \emph{IEEE International
  Symposium on Information Theory (ISIT)}, 2016, pp. 1710--1714.

\bibitem{litan}
Y.~{Li} and V.~Y.~F. {Tan}, ``Second-order asymptotics of sequential hypothesis
  testing,'' \emph{IEEE Transactions on Information Theory}, vol.~66, no.~11,
  pp. 7222--7230, 2020.

\bibitem{Marco2017}
M.~Berta, O.~Fawzi, and M.~Tomamichel, ``On variational expressions for quantum
  relative entropies,'' \emph{Letters in Mathematical Physics}, vol. 107,
  no.~12, pp. 2239--2265, 2017.

\bibitem{Wilde2013}
M.~M. Wilde, \emph{Quantum Information Theory}.\hskip 1em plus 0.5em minus
  0.4em\relax Cambridge University Press, 2013.

\bibitem{fanizza2022qusum}
M.~Fanizza, C.~Hirche, and J.~Calsamiglia, ``Qusum: quickest quantum
  change-point detection,'' \emph{arXiv preprint arXiv:2208.03265}, 2022.

\end{thebibliography}

\end{document}